\documentclass[twocolumn,showpacs,preprintnumbers,amsmath,amssymb,prl]{revtex4}

\usepackage{graphicx}
\usepackage{bm}

\usepackage{amsfonts}

\begin{document}
\title{Stable three-dimensional spatially modulated vortex solitons
 in Bose-Einstein condensates}
\author{Volodymyr M. Lashkin  }
\email{vlashkin@kinr.kiev.ua}
 \affiliation{Institute for Nuclear
Research, Pr. Nauki 47, Kiev 03680, Ukraine}

\date{\today}

\begin{abstract}

 We
present exact numerical solutions in the form of spatially
localized three-dimensional (3D) nonrotating and rotating
(azimuthon) multipole solitons in the Bose-Einstein condensate
(BEC) confined by a parabolic trap. We numerically show that the
3D azimuthon solutions exist as a continuous family parametrized
by the angular velocity (or, equivalently, the modulational
depth). By a linear stability analysis we show that 3D azimuthons
with a sufficiently large phase modulational depth can be stable.
The results are confirmed by direct numerical simulations of the
Gross-Pitaevskii equation.
\end{abstract}


\maketitle

Recently, a novel class of two-dimensional (2D) spatially
localized vortices with a spatially modulated phase, the so called
azimuthons, was introduced in Ref.~\cite{Kivshar2}. Azimuthons
represent intermediate states between the radially symmetric
vortices and nonrotating multipole solitons. In contrast to the
linear vortex phase, the phase of the azimuthon is a staircaselike
nonlinear function of the polar angle. Various kinds of azimuthons
have been shown to be stable in media with a nonlocal nonlinear
response \cite{Lopez1,Lopez2,Skupin,We2,We3}. Note, that azimuthon
solutions were found in Refs. \cite{Kivshar2,Lopez1,Lopez2} by
using an approximate (though rather accurate) variational approach
and separation of variables. The first example of exact
numerically found azimuthon solutions was presented in Ref.
\cite{We4}.

The aim of this Letter is to present exact numerical solutions in
the form of spatially localized nonrotating and rotating
(azimuthon) multipole solitons in the three-dimensional (3D) BEC
confined by a parabolic trap. We numerically show that the
azimuthon solutions exist as a continuous family parametrized by
the angular velocity or, equivalently, by the parameter which
determines the phase modulational depth. To our knowledge, this is
the first time that solutions with such a nontrivial (3D
azimuthon) topology can be found with very high (up to machine
precision) accuracy. Moreover, by means of a linear stability
analysis, we investigate the stability of these structures and
show that rotating 3D dipole solitons (azimuthons with two
intensity peaks) are stable provided that the number of atoms is
small enough and the phase modulational depth is large enough.
Thus, we present the first example of stable  rotating dipole
solitons in media with local nonlinearity. The presence of the
trapping potential plays a crucial role in stabilizing the
solitons. The results were confirmed by direct numerical
simulations of the 3D Gross-Pitaevskii equation (GPE).

 We consider a condensate at zero temperature confined in an
axisymmetric  harmonic trap. The dynamics of the condensate is
described by the normalized GPE for the wave function $\psi$
\begin{equation}
\label{main} i\frac{\partial\psi}{\partial
t}=-\Delta\psi+(x^{2}+y^{2}+\Omega^{2}z^{2})\psi-\sigma|\psi|^{2}\psi,
\end{equation}
where appropriate dimensionless units are used, $\sigma=\pm1$,
where the $+(-)$ sign corresponds to attractive (repulsive)
contact interaction.

Equation (\ref{main}) conserves the 3D norm (the normalized number
of particles) $N=\int |\psi|^{2}d\mathbf{r}$, $z$-component of the
angular momentum $M_{z}= \mathrm{Im}\,\int
\left[\psi^{\ast}(\mathbf{r}\times\nabla_{\perp}\psi)\right]_{z}
d\mathbf{r}$, and energy
\begin{equation}
\label{E} E=\int
\left\{|\nabla\psi|^{2}+(x^{2}+y^{2}+\Omega^{2}z^{2})|\psi|^{2}
-\frac{\sigma}{2}|\psi|^{4}\right\}d\mathbf{r}.
\end{equation}

We will seek solutions of Eq. (\ref{main}) which are stationary in
the frame rotating with the angular velocity $\omega$. In
cylindrical coordinates $(r,z,\varphi)$, such solutions of the
form
\begin{equation}
\psi(r,z,\varphi,t)=\Phi(r,z,\varphi-\omega t)\exp(-i\mu t),
\end{equation}
where $\mu$ is the chemical potential in the rotating frame,
satisfy the equation
\begin{gather}
\frac{\partial^{2}\Phi}{\partial
r^{2}}+\frac{1}{r}\frac{\partial\Phi}{\partial
r}+\frac{1}{r^{2}}\frac{\partial^{2}\Phi}{\partial\theta^{2}}
-i\omega\frac{\partial\Phi}{\partial\theta}+\mu\Phi \nonumber
\\ -(r^{2}+\Omega^{2}z^{2})\Phi +\sigma|\Phi|^{2}\Phi=0,
\label{main2}
\end{gather}
where $\theta=\varphi-\omega t$, and resolve the variational
problem $\delta S=0$ for the functional $S=E-\mu N-\omega M_{z}$.
The chemical potential in the laboratory frame $\mu_{l}$ is
related to the chemical potential in the rotating frame $\mu$ by
\begin{equation}
\label{mul} \mu_{l}=\mu-\omega M_{z}/N.
 \end{equation}

\begin{figure}
\includegraphics[width=3.4in]{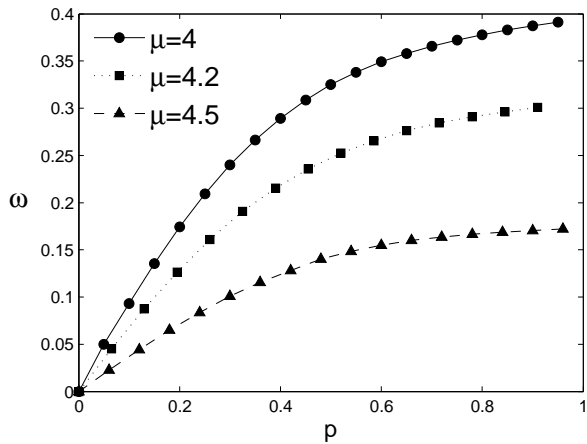}
\caption{\label{fig1} Dependence of the angular velocity $\omega$
of a dipole 3D azimuthon on the modulational depth $p$ for three
values of $\mu$ and $\sigma=1$ (attractive interaction). }
\end{figure}

\begin{figure}
\includegraphics[width=3.4in]{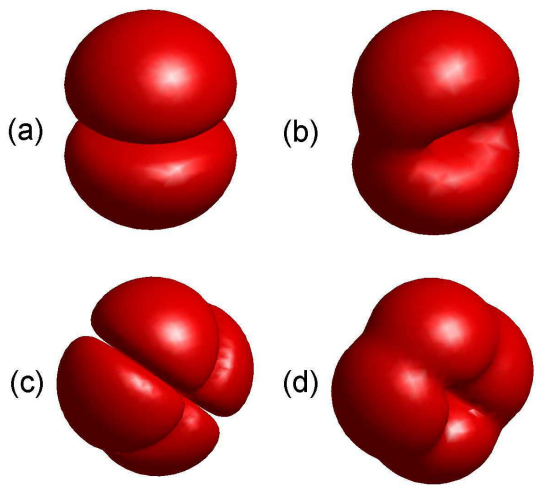}
\caption{\label{fig2} (color online) Numerically found stationary
localized (a), (c) nonrotating ($\omega=0$) and (b), (d) rotating
multipole 3D solutions of Eq. (\ref{main3}) with $\sigma=1$
(attractive interaction), $\Omega=1$, and $\mu=4.5$: (a) Dipole;
(b) azimuthon with two intensity peaks and $\omega=0.13$,
$p=0.43$; (c) quadrupole; (d) azimuthon with four intensity peaks
and $\omega=0.3$, $p=0.22$. Shown are the isosurfaces of the
soliton amplitude.}
\end{figure}

In what follows, we restrict ourselves to the case of a
spherically symmetric trap ($\Omega=1$) and attractive interaction
($\sigma=1$). Two-dimensional case corresponding $\Omega\gg 1$
("pan-cake configuration") was considered in Ref. \cite{We4}.
Equation (\ref{main2}) was rewritten in Cartesian coordinates
\begin{gather}
\mu\Phi+\Delta\Phi+(x^{2}+y^{2}+\Omega^{2}z^{2})\Phi \nonumber
\\-i\omega\left(x\frac{\partial}{\partial
y}-y\frac{\partial}{\partial x}\right)\Phi +\sigma|\Phi|^{2}\Phi=0
. \label{main3}
\end{gather}
and solved numerically with zero boundary conditions. The
numerical method that we use to find solutions of Eq.
(\ref{main3})
 was first introduced by Petviashvili \cite{Petviashvili1} (see
also Ref. \cite{Petviashvili2}) and more recently in Refs.
\cite{Pelinovsky1,Ablowitz,Fibich,Pelinovsky2,Lakoba1}.
Petviashvili employed a stabilizing factor to suppress divergence
of the iteration procedure (or convergence to zero solution).
Petviashvili and the authors of Refs.
\cite{Pelinovsky1,Ablowitz,Fibich,Pelinovsky2} considered periodic
boundary conditions and worked in the spectral Fourier
representation. We work in physical space and use zero boundary
conditions so that the method should be modified. We describe it
in the following way. Equation (\ref{main3}) can be written as
$L\Phi=N(\Phi)$, where $L\Phi$ stands for the linear part of the
equation and $N(\Phi)$ is the nonlinear term. At each $n$th stage
of the iteration
 we solve the corresponding linear problem
 and the iteration procedure is
\begin{equation}
\label{iter} \Phi_{n+1}=s^{\lambda}L^{-1}N(\Phi_{n}),
\end{equation}
where the stabilizing factor
\begin{equation}
s=\frac{(\Phi^{\ast}_{n},\Phi_{n})}{(\Phi^{\ast}_{n},L^{-1}N(\Phi_{n}))},
\end{equation}
where $(u,v)\equiv \int uv \,dxdy$, and for the cubic nonlinearity
$\lambda=3/2$. Note, that one can also take (sometimes it yields
faster convergence)
\begin{equation}
s=\frac{(\Phi^{\ast}_{n},\Phi_{n})}{(L^{-1}N(\Phi_{n}^{\ast}),L^{-1}N(\Phi_{n}))},
\end{equation}
with $\lambda=3/4$ in Eq. (\ref{iter}). It can be seen that the
right-hand side of Eq. (\ref{iter}) has homogeneity zero with
respect to $\Phi$, and this, indeed, prevents the aforementioned
divergence. The convergence can be monitored using the value
$|s-1|$, which should approach zero.

As an initial guess we use
\begin{equation}
\label{trial}
\Phi(r,z,\theta)=Ar^{m}e^{-r^{2}/a^{2}-z^{2}/b^{2}}(\cos
m\theta+i\,p_{0}\sin m\theta),
\end{equation}
on a Cartesian grid, where $m$ is an integer. The parameter
$p_{0}$ determines the modulation depth of the soliton intensity
and without loss of generality we can assume $0\leqslant
|p_{0}|\leqslant 1$. The ansatz Eq. (\ref{trial}) describes
topology of the 3D azimuthon solutions. Note that the case
$p_{0}=0$ corresponds to nonrotating multipole 3D solitons (e. g.
$m=1$ to a dipole, $m=2$ to a quadrupole etc.), while the opposite
case $|p_{0}|=1$ corresponds to the radially symmetric 3D vortices
with the topological charge $m$. A detailed study, including a
linear stability analysis, of 3D vortices in  BEC with attractive
interactions and a parabolic trap was performed in Refs.
\cite{Malomed1,Malomed2}.The intermediate case $0<|p_{0}|<1$
corresponds to the rotating 3D azimuthons.

Note, that Eq. (\ref{main3}) with $\sigma=1$ has localized
solutions provided that $\mu_{l}<4+\Omega$ \cite{Malomed1}, where
$\mu_{l}$ is the chemical potential in the laboratory frame. The
fundamental solitons of Eq. (\ref{main3}) with $\sigma=1$ and
$\Omega=1$ exist only if $\mu_{l}<3$. Moreover, the radially
symmetric 3D vortices with $m=1$ are stable if $3.72<\mu_{l}<5$
(for $\Omega=1$) \cite{Malomed1,Malomed2}, so that in the
following we consider only the region $3.8<\mu<5$.

The rate of convergence does not depend on the choice of $A$, $a$,
and $p_{0}$. In all runs we used $A=1$, $a=b=1$, and $p_{0}$
between $0$ and $1$. During the iterations, the parameter $p$
(modulational depth), which is similar to the parameter $p_{0}$ in
Eq. (\ref{trial}), can be introduced (after eliminating the
constant phase factor in $\Phi_{n}$) in the following way:
\begin{equation}
p_{n}=\max|\mathrm{Im}\,\Phi_{n}|/\max|\mathrm{Re}\,\Phi_{n}|.
\end{equation}
We start with the case $m=1$. The iteration procedure
monotonically converges to a solution that consists of two
dipole-shaped structures in the real and imaginary parts of $\Phi$
(dipole azimuthon, or, azimuthon with two intensity peaks). If the
amplitudes of these parts are equal (i. e. $p=1$) we get the
radially symmetric 3D vortex with the topological charge $m=1$. If
these amplitudes are different ($0<p<1$), we have a rotating
azimuthon; if one of the amplitudes is zero ($p=0$), we have the
nonrotating 3D dipole soliton. The final value of $p$ depends only
(for the fixed $\mu$) on the rotational velocity $\omega$ and does
not depend on the initial guess of $p$ (for example,  the initial
condition with $\omega=0$ and $p=1$ converges to the nonrotating
dipole with $p=0$). Typically we used a spatial grid with a
resolution $100^{3}$.

The progressive iterations in the relaxation method were
terminated when the norm of the residual fell below
$10^{-7}-10^{-12}$ (with the corresponding
$|s-1|<10^{-8}-10^{-13}$). Under this, the relative differences
between the successive amplitudes $|A_{n+1}-A_{n}|/A_{n}$, where
$A_{n}=\max |\Phi_{n}|$, and modulational depths
$|p_{n+1}-p_{n}|/p_{n}$ were less than $10^{-6}$ and $10^{-5}$
respectively. Typically, from $9$ (for solutions with small $p$)
to $160$ (for solutions with $p$ close to $1$) iterations were
required for convergence. Increasing the number of iterations and
grid points, we were able to find the solutions with accuracy up
to machine precision.

\begin{figure}
\includegraphics[width=3.4in]{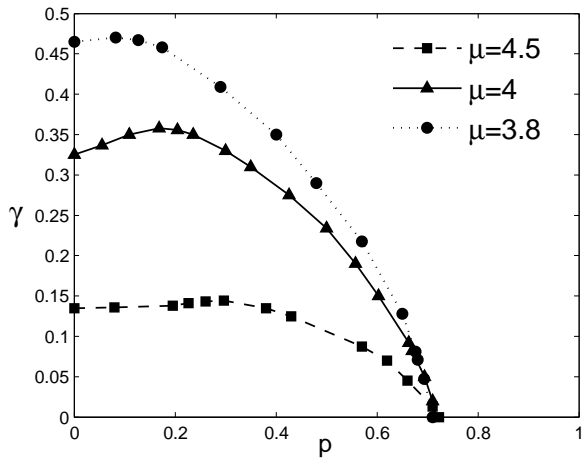}
\caption{\label{fig3} The growth rates $\gamma$ as functions of
the phase modulational depth $p$ for  different $\mu$.}
\end{figure}

By changing $\omega$ one can find the azimuthon solution with
arbitrary $p$ and, thus, there exists  a continuous family of
azimuthons parametrized by the parameter $p$. The dependence of
the angular velocity $\omega$ of the azimuthon with two intensity
peaks on the modulational depth $p$ for several different $\mu$
and $\sigma=1$ (attractive interaction) is presented in
Fig.~\ref{fig1}. In Figs. ~\ref{fig2}(a) and (b) we demonstrate
two numerically found examples of the azimuthons (nonrotating
dipole and rotating azimuthon with two intensity peaks) for the
model described by Eq. (\ref{main}). Similar solutions can be
found for $\sigma=-1$ (repulsive interaction) and $\mu_{l}>5$.

The vortex solitons $(p =1)$ have the maximum angular momentum
$M_{z}$, while the dipole solitons $(p =0)$ have zero angular
momentum. Structures with the intermediate values of $p$ carry out
the nonzero angular momentum, and this leads to soliton rotation.

Changing topology of the initial guess one can find high-order
azimuthon solutions. We took Eqs. (\ref{trial}) with $m=2$. The
iteration procedure began to converge monotonically to a solution
that consists of two quadrupole-shaped structures in the real and
imaginary parts of $\Phi$. If the amplitudes of these parts are
equal ($p=1$) we get the radially symmetric vortex with the
topological charge $m=2$. In the intermediate case ($0\leqslant
p<1$), we have the azimuthon with four intensity peaks. The
convergence was controlled by stopping the iteration when the
value $|s-1|$ began to increase. This indicates that the iteration
procedure jumps off the solution corresponding high-order
azimuthon (and begins to converge to the azimuthon with two
intensity peaks). Nevertheless, the high-order solutions can be
found with a rather high accuracy: typically, one can reach
$|s-1|\sim 10^{-3}$. Then, the obtained solution can be used as an
initial condition  in the Yang-Lakoba iterative procedure
\cite{Lakoba2} which belongs to a family of universally-convergent
iterative methods and can converge to any nonfundamental (i. e.
high-order) solution of a given equation provided that the initial
condition is \textit{sufficiently close} to that solution. Thus,
we were able to find high-order azimuthons of Eq. (\ref{main})
with very high accuracy. The nonrotating quadrupole and rotating
azimuthon with four intensity peaks are presented in Figs.
~\ref{fig2}(c) and (d).

The physical relevance of any stationary solution, of course,
depends on whether it is stable. To study the stability of the
stationary solutions, we represent the wave function in the form
\begin{equation}
\label{per}
\psi(\mathbf{r},t)=[\varphi_{0}(\mathbf{r})+\varepsilon(\mathbf{r},t)]e^{-i\mu
t},
\end{equation}
where the stationary solution $\varphi_{0}$ is perturbed by a
small perturbation $\varepsilon$. In the absence of the radial
symmetry, the corresponding eigenvalue problem (i. e. after
linearizing and taking $\varepsilon(\mathbf{r},t)\sim
\nu(\mathbf{r})\exp(\gamma t)$) on a $N^{3}$ spatial grid implies
a $2N^{3}\times 2N^{3}$ complex nonsymmetric matrix and, for
reasonable $N$ (say, $N>100$), represents extremely difficult
task. Instead, after inserting Eq. (\ref{per}) into Eq.
(\ref{main}), we solved the Cauchy problem for the linearized
equation
\begin{gather}
i\frac{\partial\varepsilon}{\partial
t}+\mu_{l}\varepsilon+\Delta\varepsilon-(x^{2}+y^{2}+\Omega^{2}z^{2})\varepsilon
\nonumber \\ +
2|\varphi_{0}|^{2}\varepsilon+\varphi_{0}^{2}\varepsilon^{\ast}=0
\end{gather}
with some initial small perturbations $\varepsilon$. The final
results are not sensitive to the specific form of
$\varepsilon(\mathbf{r},0)$. The chemical potential in the
laboratory frame $\mu_{l}$ is determined from Eq. (\ref{mul}). If
the dynamics is unstable, the corresponding solutions
$\varepsilon(\mathbf{r},t)$, undergoing, generally speaking,
oscillations, grow exponentially in time and an estimate for the
growth rate $\gamma$ can be written as
\begin{equation}
\gamma=\frac{1}{2\Delta t}\ln\left\{\frac{P(t+\Delta
t)}{P(t)}\right\},
\end{equation}
where $P(t)=\int |\varepsilon|^{2}d\mathbf{r}$, $t$ and $\Delta t$
are assumed to be large enough.

\begin{figure}
\includegraphics[width=3.4in]{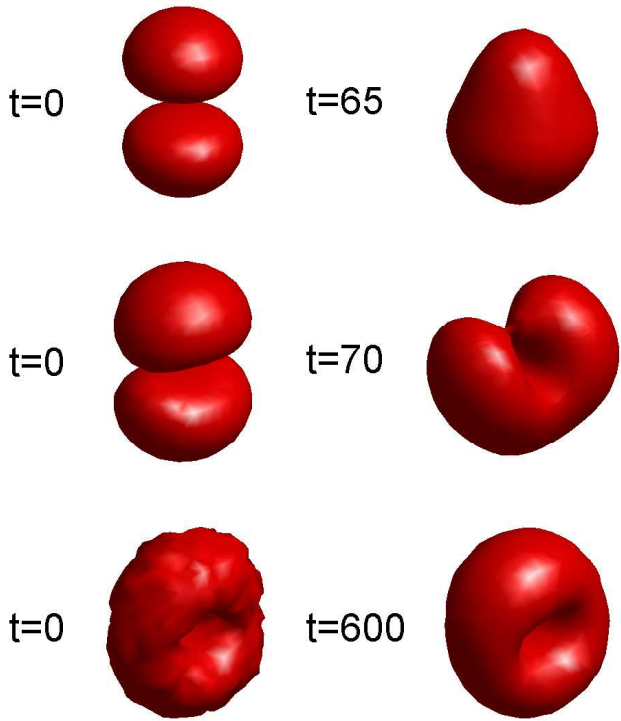}
\caption{\label{fig4} (color online) The top row: unstable
evolution of the 3D nonrotating ($\omega=0$,$p=0$) dipole with
$\mu=4.2$. Initial state of the dipole is unperturbed. The middle
row: unstable evolution of the 3D azimuthon with two intensity
peaks and $\omega=0.18$, $p=0.3$, $\mu=4.2$. The initial state is
perturbed by the white noise with the parameter $\epsilon=0.005$.
The bottom row: stable dynamics of the 3D azimuthon with two
intensity peaks and $\omega=0.285$, $p=0.73$, $\mu=4.2$. The
initial state is perturbed by the white noise with the parameter
$\epsilon=0.08$.}
\end{figure}

In Fig. \ref{fig3} we plot the growth rates $\gamma$ as functions
of the modulational depth $p$ for the azimuthons with two
intensity peaks and several different values  $\mu$ of the
chemical potential in the rotating frame. The linear stability
analysis shows that for solutions with $p\lesssim 0.7$, the growth
rate of perturbations $\gamma\neq 0$ for all $\mu$ so that there
is no stability region. In particular, all nonrotating dipoles are
unstable. The growth rate $\gamma$ decreases as $\mu$  increases
and can be very small if $\mu$ is close to $5$. The situation,
however, changes when the modulational depth exceeds the critical
value $p_{cr}\sim 0.7$. In this case, the growth rate of
perturbations $\gamma$ falls to zero, the stability window appears
and the azimuthons with $p\gtrsim 0.7$ are stable. The critical
value $p_{cr}$ varies very slightly with changing the chemical
potential $\mu$. All high-order azimuthon solutions turn out to be
unstable (for the radially symmetric vortices with $p=1$ and the
topological charge $m\geq 2$ it was shown in
Ref.~\cite{Malomed1,Malomed2}).

To verify the results of the linear analysis, we solved
numerically the dynamical equation (\ref{main}) initialized with
our computed solutions with added Gaussian noise. The initial
condition was taken in the form $\varphi_{0}[1+\epsilon
\xi(\mathbf{r})]$, where $\varphi_{0}(\mathbf{r})$ is the
numerically calculated solution, $\xi(\mathbf{r})$ is the white
gaussian noise with variance $\sigma^{2}=1$ and the parameter of
perturbation $\epsilon=0.005 \div 0.1$. The unstable dynamics of
the nonrotating dipole ($p=0$) with $\mu=4.2$ is illustrated in
the top row of Fig.\ref{fig4}. In the middle row we present
unstable evolution of the rotating azimuthon with two intensity
peaks and $\mu=4.2$, $\omega=0.18$, $p=0.3$. A slight stochastic
perturbation with $\epsilon=0.005$ was applied at $t=0$. The
azimuthon lost its shape after one rotational period. Stable
evolution of the azimuthon with $\mu=4.2$, $\omega=0.285$ and
$p=0.73$ (i. e. in the region of stability) is shown in the bottom
row of Fig.\ref{fig4}. The initial state of the azimuthon is
perturbed by a rather strong noise ($\epsilon=0.08$). The period
of rotation of the azimuthon is $T=2\pi/\omega\sim 22$. The
azimuthon cleans up itself from the noise and survives over many
dozens of the rotational periods.

In the present paper we restricted ourselves to the case of a
spherically symmetric trap ($\Omega=1$). A linear stabilty of the
radially symmetric 3D vortices (i. e. azimuthons with $p=1$ in our
case) was performed for different $\Omega$ in Ref.
\cite{Malomed2}. By analogy with the results of Ref.
\cite{Malomed2}, for the case $\Omega\ll 1$ ("cigar-like
configuration") one could expect that the quasi-1D azimuthon
solitons with $p<1$ are tightly confined in the corresponding
cigar-like trap, which suppresses their destabilization. On the
other hand, the case $\Omega \gg 1$ ("pan-cake configuration")
implies nearly 2D azimuthon solitons, with the largest energy,
which are most prone to instabilty.

In conclusion, we have presented exact numerical solutions in the
form of spatially localized 3D nonrotating and rotating
(azimuthon) multipole solitons in the three-dimensional BEC
confined by a parabolic trap. We have shown that the 3D azimuthon
solutions exist as a continuous family parametrized by the angular
velocity (or, equivalently, the phase modulational depth). By a
linear stability analysis we have shown that 3D azimuthons with a
sufficiently large modulational depth can be stable. The results
are confirmed by direct numerical simulations of the
Gross-Pitaevskii equation.

The author thanks A. I. Yakimenko and Yu. A. Zaliznyak  for
discussions.

\end{document}